# Coefficient of restitution: Derivation of Newton's Experimental Law from general energy considerations

Avi Marchewka

8 Gali Tchelet St., Herzliya, Israel

avi.marchewka@gmail.com

## Abstract

In order to describe the velocity of two bodies after they collide, Newton developed a phenomenological equation known as "Newton's Experimental Law" (NEL). In this way, he was able to practically bypass the complication involving the details of the force that occurs during the collision of the two bodies. Today, we use NEL together with momentum conservation to predict each body's velocity after collision. This, indeed, avoids the complication of knowing the forces involved in the collision, making NEL very useful. Whereas in Newton's days the quantity of kinetic energy was not known, today it is a basic quantity that is in use. In this paper we will use the loss (or gain) of kinetic energy in a collision to show how NEL can be derived.

Keywords: Newton's Experimental Law, Coefficient of restitution, Galilean invariant ,center of masses

## 1. Introduction

A typical collision scenario is presented in Figure 1 (a and b) below. Two masses, $m_1, m_2$ are shown before (Fig 1.a), and after (Fig 1.b) a collision. The masses' velocities are $v_{i,1}, v_{i,2}$ respectively initially (Fig 1.a), and $v_{f,1}, v_{f,2}$ respectively after collision (Fig 1.b).

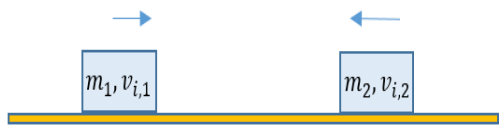

Figure 1(a) Before the two masses collide

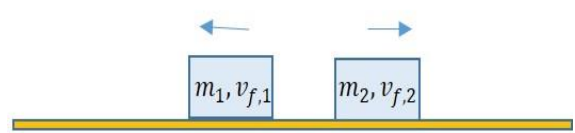

Figure 1 (b) After the two masses collide

NEL [1-4] gives the ratio between the bodies' different velocities before the collision $v_{i,1} - v_{i,2}$, after collision $v_{f,1} - v_{f,2}$ This ratio is defined by an experimental coefficient known as the "Coefficient of Restitution" (COR) denoted by $e$. Specifically,

$$e = \frac{v_{i,1} - v_{i,2}}{v_{f,2} - v_{f,1}} \qquad\qquad (1.1)$$

The left side of the equation (1.1), COR., is an *experimental* constant, dependent mainly on the material that the bodies are made of, but weakly dependent on other properties involved in the collision; in particular, the bodies' shape, temperature, and most importantly, the velocity of the two bodies [1]. This enables us to treat a large number of collisions by taking $e$ as approximat constant. COR

There are many aspects and details to the behaviour of $e$ [2], and we will not deal with all of them here. Note, however, that it is not hard to imagine cases, such as if one of the bodies gets broken or chipped during the collision. In such cases, since the bodies are not the same after the collision, clearly NEL stops being relevant for describing the physics of the collision, and the coefficient $e$ is not defined.

Historically, Newton introduced his formula in 1687 [3], one hundred and fifty years before the concept of kinetic energy as we use it today was given by Gaspard-Gustave Coriolis in 1829 [5].

An important property of COR is that it is a Galileo invariant. The difference between the two bodies' velocities is a Galileo invariant, and thus the ratio of two Galileo invariants is also a Galileo invariant. According to (1.1), it follows that COR., $e$, is a Galileo invariant [10]. This is an important property when dealing with the significance of NEL. and will be discussed below.

Nowadays, NEL appears in various areas and different contexts. It appears in most, if not all, college-level mechanics textbooks. Typically, it is used in two cases: $e = 1$, which describes the elastic collision where the kinetic energy is conserved at the collision so that $v_{i,1} - v_{i,2} = v_{f,2} - v_{f,1}$. And, $e = 0$ the complete inelastic collision, where the two colliding bodies travel together as a single extended body after the collision, $v_{f,1} = v_{f,2}$

Other values of COR., $0 < e < 1$ are evident from everyday phenomena, such as a bouncing ball, a colliding car, and so forth. In addition, it is often used in class demonstrations, as a popular topic for discussion [6], and in models [7]. It is also applied in material engineering [9], and mechanical engineering [8]. In a major part of this paper we focus on the case where $0 \le e \le 1$. The case $e > 1$ also occurs and is discussed at the end of the paper.

The main results of this paper are given in six steps (3.1-3.6) of part 3.

## 2. An experimental demonstration of NEL: bouncing a ball on the floor

One easy way to demonstrate NEL is by measuring the bounce-heights of a bouncing ball [11]. Consider dropping a spin-less ball from an initial height of $h_0$ with initial velocity $v_x, v_y$, bouncing on a flat floor with $0 < e < 1$. The ball's height decreases after each bounce. The ratio **as measured**, between two consecutive ball peaks is a constant. Denoting the sequence of the ball's peak by $q$, $q = 0$ is the initial peak, $q = 1$ the peak after the first bounce, et cetera.

The ball's bounce-height measurements behave according to,

$$\frac{h_1}{h_0} = \frac{h_2}{h_1} = \cdots = \frac{h_q}{h_{q-1}} = e^2 \tag{1.2}$$

According to the energy consideration of the decreasing bounce-heights (where at all times the floor's velocity is zero), we can find the change in velocity due to the ball's impact on the floor. Denoting the sequence of the ball hits on the floor by $r$, $r = 0$ is the first time the ball hits the floor, $r = 1$ is the second time the ball hits the floor, et cetera. That is,

$$\frac{v_{i,0}}{v_{f,0}} = \frac{v_{i,1}}{v_{f,1}} = \cdots = \frac{v_{i,r}}{v_{f,r}} = -e \tag{1.3}$$

The minus sign occurs due to the ball's change of direction at the bouncing points, and by conservation of energy $v_{f,(r-1)} = v_{i,r}$. As shown in [11], $e$ depends (albeit weakly), on the velocity of the bouncing ball when it hits the floor.

Thus, (1.2) recovered from (1.1). This demonstrates one of the many applications of NEL.

In Fig. 2, we show the heights of such a bouncing ball on the floor with $h = 2$, $v_{i,x} = v_{i,y} = 0$ and $e = 0.8$ according to (1.3). In Fig. 3. the velocity of the same bouncing ball according to (1.2) is shown.

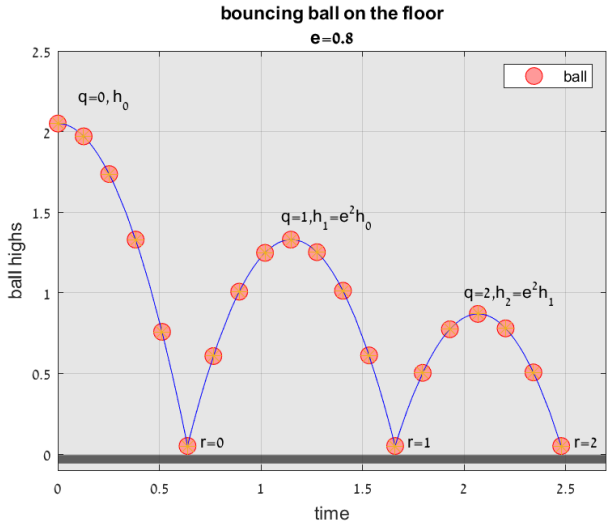

**FIGURE 2** THE DECREASING HEIGHTS OF A BOUNCING BALL AS A FUNCTION OF TIME, ACCORDING TO *(1.2)*

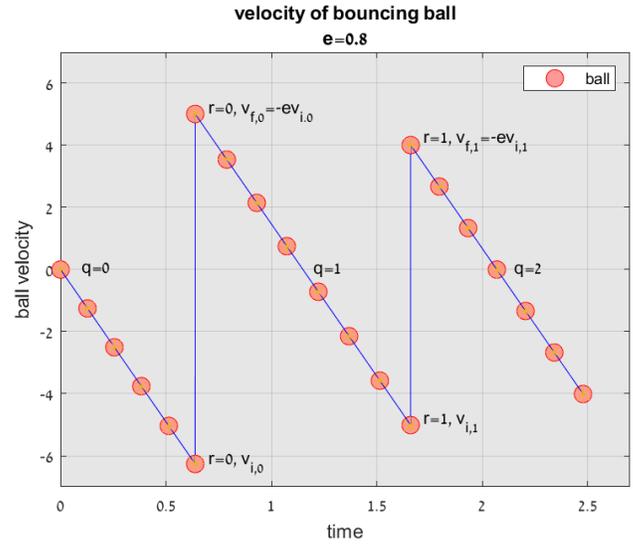

**FIGURE 3** THE VELOCITY OF A BOUNCING BALL AS A FUNCTION OF TIME ACCORDING TO *(1.3)*. NOTE THE "JUMP" AT THE VELOCITY

# 3. Derivation of Newton's Experimental Law

### 3.1 A Galilean invariant measure of kinetic energy losses in collision

As demonstrated above, it is evident from everyday experiments that colliding bodies often lose their kinetic energy during collision. So the first step will be to construct a measure of such loss of kinetic energy that will be agreed upon all Galilean Frame of Reference (FOR), i.e., a Galilean invariant measure of the loss of the kinetic energy in collision.

The kinetic energy of a system is dependent on the FOR it is measured at. Therefore, it cannot be used as a global quantity that describes the physics of the process, which involves loss of kinetic energy *per se*. Indeed, consider the kinetic energy of the two colliding bodies ( Fig. 1). In the lab system $S$, their initial and final kinetic energies are,

$$T_k^{(S)} = \frac{1}{2}\sum_{j=1}^{2} m_j v_{j,k}^2 \quad \text{for} \quad k = i, f \tag{1.4}$$

where $i$ and $f$ stand for the initial and final kinetic energy respectively, and $j = 1, 2$ stand for the first and second bodies respectively.

In $S'$ frame, with velocity $u$ relative to the lab frame system $S$, the above kinetic energies are

$$T_k^{(S')} = \frac{1}{2} \sum_{j=1}^{2} m_j \left( v_{j,k} - u \right)^2 \quad \text{for} \quad k = i, f \tag{1.5}$$

Clearly, we have $T_k^{(S)} \neq T_k^{(S')} \quad \text{for} \quad k = i, f$.

Now, instead of the kinetic energy, let us consider the difference between the initial kinetic energy of the two bodies, and the final kinetic energy after the collision of those bodies. From (1.4) in $S$ frame, this difference is

$$T_i^{(S)} - T_f^{(S)} = \frac{1}{2} \left[ m_1 \left( v_{i,1}^2 - v_{f,1}^2 \right) + m_2 \left( v_{i,2}^2 - v_{f,2}^2 \right) \right] \tag{1.6}$$

From (1.5) in $S'$, this difference is

$$T_i^{(S')} - T_f^{(S')} = \frac{1}{2}$$
$$\left[ m_1 \left( \left( v_{i,1} - u \right)^2 - \left( v_{f,1} - u \right)^2 \right) + m_2 \left( \left( v_{i,2} - u \right)^2 - \left( v_{f,2} - u \right)^2 \right) \right] \tag{1.7}$$

To see why (1.6) is equal to (1.7), we have to use Newton's 3rd law: the force acting on each body during the time of collision has equal magnitude and opposite signs. This is manifested in momentum conservation. Indeed, in each FOR momentum is conserved $p_i = p_f$. That is,

$$m_1 v_{1,i} + m_2 v_{2,i} = m_1 v_{1,f} + m_2 v_{2,f} \tag{1.8}$$

Using Equations (1.6),(1.7) and (1.8), we find

$$T_i^{(S')} - T_f^{(S')} = T_i^{(S)} - T_f^{(S)} \equiv C_0 \tag{1.9}$$

Then we see that the difference in kinetic energy, $C_0$ is invariant under all Galilean FOR.

Thus $C_0$ is our candidate to describe the collision for which all FOR agree upon.

Obviously, the kinetic energy difference is not the only possibility of Galilean invariant as a function of kinetic energy (one may find others), but it is the simplest one.

## 3.2 The upper bound of energy loss in a collision

In collision, a general question to be answered is: What is the maximum possible kinetic energy the colliding system may lose that will be agreed upon for all FOR.? Or, put another way, what is the upper limit of $C_0$ in equation (1.9)?

Consider the collision scenario in Fig 1, as viewed from difference FOR. We denoted by $S^{(j)}, j = 0, 1, 2\ldots,$ the differences FOR and by $S^0 \equiv S$ the lab FOR. Now, because the final kinetic energy is not negative in each of all those FOR. Then we have,

$$0 \le C_0 \le T_i^{(j)} \quad , \quad \forall\, j = 0, 1, 2\ldots \tag{1.10}$$

In other words, in each of the FOR the kinetic energy lost is bound by the initial kinetic energy in this FOR.

Furthermore, it follow that:

*The maximum possible kinetic energy that may be **lost** has an upper bound.*

*This upper bound energy is the initial energy of the FOR with minimal initial kinetic energy.*

That is, from (1.10)

$$0 \le C_0 \le \min\left\{ T_i^{(j)} \right\} \quad , \quad \forall\, j = 0, 1, 2\ldots \tag{1.11}$$

This suggests a particular frame of reference: *Among the infinite FOR that the collision can be described by, there is (or are) a special FOR which is characterized by having minimal total kinetic energy. It follows from* (1.11) *that the initial energy in this system is upper bound to the possible energy losses.*

We still don't know whether:

1. All this initial kinetic energy can be lost; we only know that it is bound to the energy that may be lost
2. It is a single FOR or not

We will be denoting this special FOR by $R$, and its velocity compared to the lab system $S$ by $\xi$.

## 3.3 The $R$ FOR and its properties

In order to find this (or those) $R$ FOR, consider the kinetic energy of the colliding bodies in the lab system $S$. The bodies' velocities are $v_{k,1}$ and $v_{k,2}$ respectively. From the $R$ FOR, their kinetic energy is,

$$T_k^{(R)} = \frac{1}{2} m_1 \left( v_{i,1} - \xi_k \right)^2 + \frac{1}{2} m_1 \left( v_{i,2} - \xi_k \right)^2 \tag{1.12}$$

Now, we are looking for a FOR, i.e., the velocity $\xi_k$, which minimizes (1.12) for all value of $v_{i,1}, v_{i,1}$ ($m_1, m_2$ are constant). (1.12) is parabola in the variable $\xi_k$ and thus, has a single minimum. This single minimum can be found by the extremum condition

$$\frac{d}{d\xi_k} T_k^{(R)} = -m_1 \left( v_{i,1} - \xi_k \right) - m_2 \left( v_{i,2} - \xi_k \right)$$

This linear equation with a single solution for $\xi_k$ that is,

$$\xi_k = \frac{m_1 v_{k,1} + m_2 v_{k,2}}{m_1 + m_2} \tag{1.13}$$

However, since the initial velocities $v_{i,1}, v_{i,2}$ are the velocity before the collision and $v_{f,1}, v_{f,2}$ are the velocity after the collision,

Using (1.8) and equations (1.12),(1.13) and **Error! Reference source not found.**, it follows that

$$\xi_i = \xi_f = \xi \tag{1.14}$$

We therefore found a FOR that has the property of having minimal initial kinetic energy. Furthermore, since $R$ is uniquely given by the velocity in the lab FOR $S$, it follows that $R$ is also uniquely defined.

In order to calculate the kinetic energy, initial or final, in the $R$ FOR we need the bodies' velocities,

$$v_{k,1}^{(R)} = v_{k,1} - \xi = \frac{m_2 \left( v_{k,1} - v_{k,2} \right)}{m_1 + m_2}$$

$$k = i, f \tag{1.15}$$

$$v_{k,2}^{(R)} = v_{k,2} - \xi = \frac{m_1 \left( v_{k,2} - v_{k,1} \right)}{m_1 + m_2}$$

Then, the initial and final kinetic energy in $R$ FOR. are

$$T_k^{(R)} = \frac{1}{2} m_1 \left( v_{k,1}^{(R)} \right)^2 + \frac{1}{2} m_2 \left( v_{k,2}^{(R)} \right)^2$$
$$= \frac{m_1 m_2 \left( v_{k,1} - v_{k,2} \right)^2}{2 \left( m_1 + m_2 \right)} \quad , \quad k = i, f \tag{1.16}$$

From (1.10) and (1.16), we have the upper bound of the kinetic energy in the $R$ FOR,

$$C_0 \leq T_i^{(R)} = \frac{\mu}{2} \left( v_{k,1} - v_{i,2} \right)^2 \tag{1.17}$$

where $\mu = m_1 m_2 / \left( m_1 + m_2 \right)$ is the usually the reduced mass of the system.

Note again, there is no possibility to know whether all this energy or only part of it is lost. Therefore, we found only the upper bound.

Let us summarize the properties of the system $R$ that we need for further consideration:

1. The energy of $R$ is minimal compared to all other FOR.
2. Consider the total momentum of the bodies in $R$. From (1.8) and (1.15) we have

$$P_k^{(R)} = m_1 v_{k,1}^{(R)} + m_2 v_{k,2}^{(R)} = 0 \quad , \quad k = i, f \tag{1.18}$$

Therefore the FOR $R$ has the *property* of center of mass FOR, zero total momentum, and $\mu$ is its reduced mass.

However, note that the standard derivation of the center of mass FOR starts with the defining the center of mass coordinate [4]. Accordingly in one dimension, the center of mass coordinates for two bodies is,

$$x_{c.m.} = \frac{m_1 x_1 + x_2 m_2}{m_1 + m_2}$$

In this case the velocity of the center of mass is $\dot{x}_{c.m.}$, which is $\dot{x}_{c.m.} = \xi$. Here, motivated to find the FOR that has the minimal energy, the FOR we are seeking is defined by the velocity $\xi$. This way of finding the $R$ frame may be used as an alternative and/or supplement to the ordinary definition of the center of mass frame of reference, now motivated by physical considerations.

3. Equation (1.16) shows the unique property of the total kinetic energy in $R$. Specifically, it shows that the total kinetic energy is proportional to the differences of the velocities squared, $\left(v_{k,1} - v_{k,2}\right)^2$, which is a Galilean invariant, whereas, in other FOR, the total kinetic energy is proportional to the sums of the square of each body's velocities, which is not a Galilean invariant. E.g., in the lab FOR, the total kinetic energy $\propto m_1 v_{k,1}^2 + m_2 v_{k,2}^2$.

4. Due to the mass being positive from (1.18), the velocity of the body is either opposite to the other body, or both masses have zero velocity.

5. Because, or under the assumption that the colliding bodies do not cross one another, then the sign of the relative motion of the body before the collision is opposite to the sign of relative motion after the collision. That is,

$$\mathrm{sign}\left(\mathrm{v}_{k,1}^{(R)}\right) = -\mathrm{sign}\left(\mathrm{v}_{k,2}^{(R)}\right) \quad , \quad k = i, f \tag{1.19}$$

where, and the $\mathrm{sign}$ is defined as follows

$$\mathrm{sign}\left(x\right) = \begin{cases} -1 & x < 0 \\ 0 & x = 0 \\ 1 & x > 0 \end{cases}$$

This, for example, implies that if the initial velocity is

$$v_{i,1}^{(R)} > v_{i,2}^{(R)}$$

then after the collision

$$v_{f,1}^{(R)} < v_{f,2}^{(R)}$$

### 3.4 The lower and the upper value of the kinetic energy that can be lost in collision

Previously, according to (1.17), the possible loss of energy has given. Now we turn to find the actual upper and lower values of the energy lost in collision, which arise from the physical equations given in (1.17) and (1.18).

Using the fact that in frame $R$ the total momentum of the system is zero (1.18), we rewrite the kinetic energy of the system at $R$ (1.16) as function of a single velocity. The initial kinetic energy in terms of the initial velocity is,

$$T_i^{(R)} = \frac{m_1 m_2 + m_1^2}{2m_2} \left( v_{i,1}^{(R)} \right)^2$$

$$= \frac{m_1 m_2 + m_2^2}{2m_1} \left( v_{i,2}^{(R)} \right)^2 \tag{1.20}$$

and similarly, using (1.16) and (1.18) for the final velocities, the final kinetic energy in terms of the final velocity is,

$$T_f^{(R)} = \frac{m_1 m_2 + m_1^2}{2m_2} \left( v_{f,1}^{(R)} \right)^2$$

$$= \frac{m_1 m_2 + m_2^2}{2m_1} \left( v_{f,2}^{(R)} \right)^2 \tag{1.21}$$

Note that in (1.20) and (1.21), the direction of the velocities disappears. In order to find the sign of the velocity, one can use the momentum equation (1.18). Now, since there is no additional constraint between the initial and final velocity square, (1.20) and (1.21) are independent from each other.

In order to find the lower and upper bound of the kinetic energy losses out of those equations, we first look at the possible solution of (1.21). Whereas the value of the initial energy in (1.20) is known, the final value of the velocity depends on the specification of the forces during the collision, and so the value of the final energy is given in (1.21). Indeed, there is a vast variety of forces that could occur, which result in range of final velocities. We are interested in the upper value and lower value.

The solution with the lowest value of finale energy is,

$$v_{f,1}^{(R)} = v_{f,2}^{(R)} = 0 \tag{1.22}$$

Therefore, from (1.21), the lower value of the kinetic energy $T_f^{(R)}$, is zero. It is informative to see what the meaning of (1.21) is for other FOR. By Galilean transformation of (1.21), the bodies velocities are $v_{f,1}^{(j)} = v_{f,2}^{(j)}$, where $j$ is any other FOR. This means that after the collision, the bodies travel together in all FOR . This is known as the perfect in-elastic collision, in which the bodies travel together after the collision.

To find the upper value of the final kinetic energy $T_f^{(R)}$ for the case of kinetic energy losses, we see that (1.21) has maximum value equal to the initial kinetic energy. From (1.21) and (1.20), the solution for the final maximum energy is

$$\left(v_{i,1}^{(R)}\right)^2 = \left(v_{f,1}^{(R)}\right)^2$$

$$\left(v_{i,2}^{(R)}\right)^2 = \left(v_{f,2}^{(R)}\right)^2 \tag{1.23}$$

That is, the kinetic energy is unchanged. However, as discussed (see (1.19) and the paragraph below), the final velocity changes to

$$v_{i,1}^{(R)} = -v_{f,1}^{(R)}$$

$$v_{i,2}^{(R)} = -v_{f,2}^{(R)} \tag{1.24}$$

This case is the well known 'elastic collision'.

To conclude, we have found the minimum and maximum energies that may be lost as a result of collision.

Therefore, the following inequality holds,

$$0 \le T_f^{(R)} \le T_i^{(R)} \tag{1.25}$$

Dividing (1.21) by (1.20) and with the use of (1.25), we find

$$0 \le \frac{T_f^{(R)}}{T_i^{(R)}} = \left(\frac{v_{f,1}^{(R)}}{v_{i,1}^{(R)}}\right)^2 = \left(\frac{v_{f,2}^{(R)}}{v_{i,2}^{(R)}}\right)^2 \le 1 \tag{1.26}$$

This equation converts the condition from the kinetic energy to the various velocities of the two bodies.

Note that if $v_{i,1}^{(R)} = 0$ according to (1.18), then also $v_{i,2}^{(R)} = 0$. That is, no collision in $R$. It follows, by Galilean transformation, that there is no collision in any other FOR.

## 3.5 Newton's Experimental Law for $e \le 1$

For the case of kinetic energy lost in collision, we use (1.26), defining the relation by the parameter $e_1^2$

$$\left(\frac{v_{f,1}^{(R)}}{v_{i,1}^{(R)}}\right)^2 = \left(\frac{v_{f,2}^{(R)}}{v_{i,2}^{(R)}}\right)^2 = e_1^2 \tag{1.27}$$

where from (1.26) $0 \le e_1^2 \le 1$. Taking the square root gives

$$v_{f,1}^{(R)} = -e_1 v_{i,1}^{(R)}$$
$$v_{f,2}^{(R)} = -e_1 v_{i,2}^{(R)} \tag{1.28}$$

where we chose the minus sign due to the property in (1.19). In $R$, after the collision, the velocity changes direction.

Taking the difference between equations in (1.28), we get

$$\frac{v_{i,1}^{(R)} - v_{i,2}^{(R)}}{v_{f,2}^{(R)} - v_{f,1}^{(R)}} = e_1 \tag{1.29}$$

In the lab frame, (1.29) reads

$$\frac{\left(v_{i,1} - \xi\right) - \left(v_{i,2} - \xi\right)}{\left(v_{f,2} - \xi\right) - \left(v_{f,1} - \xi\right)} = \frac{v_{i,1} - v_{i,2}}{v_{f,2} - v_{f,1}} = e_1 \tag{1.30}$$

Now, identifying $e_1 = e$, (1.30) gives NELfor loss of energy (1.1).

It is interesting to see how (1.28) is transformed into other FOR; for example, in the lab FOR those equations become

$$v_{f,1} = -e_1 v_{i,1} - \left(e + 1\right)\xi$$
$$v_{f,2} = -e_1 v_{i,2} - \left(e + 1\right)\xi \tag{1.31}$$

which, unlike (1.28), doesn't have clear insight.

### 3.6 Kinetic energy gains in collision, $e > 1$

The condition in which final kinetic energy cannot be bigger than the initial kinetic energy is not absolute. Indeed, as can be seen from (1.20) and (1.21), the final kinetic energy can exceed the initial kinetic energy. Furthermore, collisions that result in kinetic energy gain are well known. To include energy gain, one can modify (1.10) into

$$T_f^{(R)} > T_i^{(R)} \tag{1.32}$$

Using (1.20) and (1.21), we have

$$\frac{T_f^{(R)}}{T_i^{(R)}} = \left(\frac{v_{f,1}^{(R)}}{v_{i,1}^{(R)}}\right)^2 = \left(\frac{v_{f,2}^{(R)}}{v_{i,2}^{(R)}}\right)^2 > 1 \qquad (1.33)$$

Repeating the corresponding steps from (1.27)-(1.30), NEL for gain of energy is constructed (1.1) with $e \geq 0$.

# 4.Summary

NEL (1.1), is very useful. The central motivation of this paper is to provide a logical way to construct NEL. It shows that this can be done by using everyday, observable facts; that during the collision of two bodies, their kinetic energy may either be lost, gained or unchanged.

NEL is constructed in the following six steps given in section 3:

- (3.1) We showed that the difference in kinetic energy is a Galilean invariant (1.9). Hence, we concluded that this difference is a good candidate for describing the collision between two bodies.
- (3.2) The fact that maximal possibility of kinetic energy loss is at FOR with minimal initial energy (1.11) leads us to suggest that for a collision description there is a preferable and natural FOR.
- (3.3) Due to the condition of FOR with minimal initial kinetic energy, we found the velocity of the $R$ frame in respect to the lab frame (1.15), and it is unique (1.14). This **turns out** to be the center of mass FOR
- (3.4) By combining the maximum energy loss and the fact that kinetic energy is not conserved, we found the upper value of the kinetic energy lost in collision (1.26).
- (3.5) NEL was reconstructed (1.30).
- (3.6) The case of gained kinetic energy was discussed, showing that it also reconstructs NEL for gaining energy.

Acknowledgments:

I wish to thank Dr. Yiftach Navot for our fruitful discussion.